\documentclass[a4paper,11pt]{article}
\usepackage{pos}

\usepackage{csquotes}    
\usepackage{tikz}        
\usepackage{braket}      
\usepackage{placeins}    

\newcommand{\overlayText}[5]{
  \begin{tikzpicture}
    \centering
    \node at (0, 0) {\includegraphics[width=#1,keepaspectratio,origin=c]{#2}};
    \node at (#3,#4) {#5};
  \end{tikzpicture}
}

\newcommand{\Oh}{\hat{\mathcal{O}}}
\DeclareRobustCommand{\order}[1]{\mathcal{O}\left(#1\right)}

\newcommand{\tcr}[1]{\textcolor{black}{#1}}

\newcommand{\rb}[1]{\!\left(#1\right)}
\newcommand{\sq}[1]{\left[#1\right]}
\newcommand{\expe}[1]{\text{e}^{#1}}
\DeclareRobustCommand{\eqnr}[1]{Eq.~$\left(\ref{#1}\right)$}

\newcommand{\Fig}[1]{Fig.~\ref{#1}}
\newcommand{\Tab}[1]{Table \ref{#1}}

\newcommand{\Refl}[1]{Ref.~\cite{#1}}    
\newcommand{\Refltwo}[2]{Refs.~\cite{#1,#2}}    

\title{Anisotropic excited bottomonia from a basis of smeared operators}

\author*[a]{Ryan Bignell}
\author[b]{Gert Aarts}
\author[b]{Chris Allton}
\author[b]{M.~Naeem Anwar}
\author[b]{Timothy J. Burns}
\author[c]{Rachel Horohan D'arcy}
\author[d]{Benjamin J\"ager}
\author[e]{Seyong Kim}
\author[f]{Maria Paola Lombardo}
\author[a]{Sin\'ead Ryan}
\author[a,c]{Jon-Ivar Skullerud}
\author[b]{Antonio Smecca}

\affiliation[a]{School of Mathematics \& Hamilton Mathematics Institute,
  Trinity College, Dublin, Ireland}

\affiliation[b]{Department of Physics,
   Swansea University, Swansea, SA2 8PP, United Kingdom}

\affiliation[c]{Department of Physics
  National University of Ireland Maynooth, County Kildare, Ireland}

\affiliation[d]{CP3-Origins \& Danish IAS, Department of Mathematics and Computer Science,\\
  University of Southern Denmark, 5230, Odense M, Denmark}
\affiliation[e]{Department of Physics,
  Sejong University, Seoul 143-747, Korea}

\affiliation[f]{INFN,
  Sezione di Firenze, 50019 Sesto Fiorention (FI), Italy}

\emailAdd{bignellr@tcd.ie}
\emailAdd{\{g.aarts,c.allton,m.n.anwar,t.burns,antonio.smecca\}@swansea.ac.uk}
\emailAdd{rachel.horohandarcy.2018@mumail.ie}
\emailAdd{jaeger@imada.sdu.dk}
\emailAdd{skim@sejong.ac.kr}
\emailAdd{mariapaola.lombardo@lnf.infn.it}
\emailAdd{ryan@maths.tcd.ie}
\emailAdd{jonivar@thphys.nuim.ie}

\abstract{
Bottomonia play a crucial role in our understanding of the quark gluon plasma. We present lattice non-relativistic QCD calculations of bottomonia at temperatures in the range $T\in\left[47,\,380\right]$ MeV using the \textsc{Fastsum} Generation 2L anisotropic $N_f = 2 + 1$ ensembles. The use of a basis of smeared operators allows the extraction of excited-state masses at zero temperature and an investigation of their thermal properties at non-zero temperature. We find that the ground state signal is substantially improved by this variational approach at finite temperature. We also apply the time-derivative moments approach to the projected or optimal correlation functions at finite temperature.
}

\FullConference{The 41st International Symposium on Lattice Field Theory (LATTICE2024)\\
 28 July - 3 August 2024\\
Liverpool, UK\\}

\begin{document}
\maketitle
\section{Introduction}
The role and behaviour of hadrons under extreme temperatures are key questions in our understanding of quantum chromodynamics (QCD). The temperature increase causes the confining world to transition to a deconfined quark-gluon plasma (QGP) which has deconfined light degrees of freedom and chiral symmetry restored. While the light hadrons become deconfined in the vicinity of the crossover transition, thermal modifications to the heavy (charm and bottom) hadrons, which may survive in the QGP, are a good probe of the hot medium created in heavy-ion collisions~\cite{Brambilla:2010cs,Aarts:2016hap,Zhao:2020jqu}.
\par
The properties of hadrons are encoded in the spectral functions $\rho\rb{\omega}$ which are related to the Euclidean correlators computable by lattice QCD by
\begin{align}
  G\rb{\tau;T} = \int_0^{\infty}\,\frac{d\,\omega}{2\,\pi}\,\rho\rb{\omega;T}\,K\rb{\tau,\omega;T},
\end{align}
for some temperature $T$, where $K\rb{\tau,\omega;T}$ is some known kernel function. Determining $\rho\rb{\omega}$ from the Euclidean correlator is an ill-posed problem which many methods attempt to handle~\cite{Hansen:2017mnd,Hansen:2019idp,Spriggs:2021dsb,Rothkopf:2022fyo,Bennett:2024cqv,Jay:2025dzl}. For the non-relativistic QCD (NRQCD)~\cite{Lepage:1992tx,HPQCD:2011qwj,Aarts:2010ek,Aarts:2014cda} approach to the simulation of the bottom quark employed in this work, the kernel is $\exp\rb{-\omega\,\tau}$ and so this determination amounts to an inverse Laplace transform.
\par
In this study, we will examine the fate of bottomonia as the temperature increases using the \textsc{Fastsum} anisotropic Generation 2L ensembles. We will focus in particular on the low-lying $\Upsilon$ (S wave) and the $\chi_{b1}$ (P wave) states.
\section{Ensembles}
\label{sec:LQCD:Ens}
\begin{table}[b]
  \centering
    \caption{\textsc{Fastsum} Generation 2L ensembles used in this work. The lattice size is $32^3 \times N_\tau$, with temperature $T = 1/\rb{a_\tau N_\tau}$. We use $\sim 1000$ configurations and up to $N_\tau$ Coulomb gauge fixed wall-sources. The estimate for $T_{c}$ comes from an analysis of the renormalised chiral condensate and equals $T_{c} = 167(2)(1)$ MeV~\cite{Aarts:2020vyb,Aarts:2022krz}. Full details of these ensembles may be found in Refs.~\cite{Aarts:2020vyb,Aarts:2022krz}.
    }

  \begin{tabular}{r|rrrrr||rrrrrr}
  $N_\tau$ & 128 & 64 & 56 & 48 & 40 & 36 & 32 & 28 & 24 & 20 & 16\\ \hline
  $T\,\rb{\text{MeV}}$ & 47 & 95 & 109 & 127 & 152 & 169 & 190 & 217 & 253 & 304 & 380 \\
\end{tabular}
\label{tab:ensembles}
\end{table}
The thermal ensembles of the \textsc{Fastsum} collaboration~\cite{Aarts:2020vyb,Aarts:2022krz,aarts_2025_10636046} are used in this study; here the temperature is varied via the \enquote{fixed-scale} approach on anisotropic lattices.
\par
The renormalised anisotropy is $\xi \equiv a_s/a_\tau = 3.453(6)$ \cite{Dudek:2012gj,Aarts:2020vyb}. The lattice action follows that of the Hadron Spectrum Collaboration~\cite{Edwards:2008ja} and is a Symanzik-improved~\cite{Symanzik:1983dc,Symanzik:1983gh} anisotropic gauge action with tree-level mean-field coefficients and a mean-field-improved Wilson-clover~\cite{Sheikholeslami:1985ij,Zanotti:2004qn} fermion action with stout-smeared links~\cite{Morningstar:2003gk}. Full details of the action and parameter values can be found in \Refl{Aarts:2020vyb}. The spatial lattice spacing is $a_s = 0.11208\rb{31}$ fm~\cite{Wilson:2019wfr} giving a pion mass $m_\pi=239(1)$ however we use the (earlier) scale setting from \Refl{Wilson:2015dqa}, $a_\tau=0.0330(2)$ fm in this study. The strange quark has been approximately tuned to its physical value via the tuning of the light and strange pseudoscalar masses~\cite{HadronSpectrum:2008xlg,HadronSpectrum:2012gic,Cheung:2016bym}.
\par
The ensembles are generated using a fixed-scale approach, such that the temperature is varied by changing $N_\tau$, as $T=1/\rb{a_\tau N_\tau}$. A summary of the ensembles is given in \Tab{tab:ensembles}. There are five ensembles below the pseudocritical temperature $T_{c} = 167(2)(1)$, one close to $T_{c}$ and five above $T_{c}$. The estimate for $T_{c}$ comes from an analysis of the renormalised chiral condensate~\cite{Aarts:2020vyb}. Note that here we have used the updated lattice spacing of \Refl{Wilson:2019wfr}, which has been implemented in our analysis in Ref.~\cite{Aarts:2022krz}.
\section{Operator Basis}
In order to extract the excited states of the spectrum, it is common practice in lattice field theory to consider a basis of operators rather than a single operator. The resulting matrix of correlators can be used to solve a generalised eigenvalue problem (GEVP)~\cite{Michael:1985ne,Luscher:1990ck,UKQCD:1993gym,Melnitchouk:2002eg,Burch:2006cc,Blossier:2009kd,Stokes:2018emx} which diagonalises the matrix such that a new set of operators are produced that each couple to an individual state. In practice, it is essential to ensure that the operator basis is sufficiently broad as to couple to the states of interest and nearby states.
\par
To briefly present the GEVP approach, first a matrix of correlators
\begin{align}
  G_{ij}\rb{\tau} \propto \braket{\Omega|\Oh_i\rb{\tau}\,\Oh_j^\dagger\rb{0}|\Omega},
\end{align}
is normalised and symmetrised. Here $i,\,j$ represent the operators used and $\Omega$ the QCD vacuum state. Then the generalised eigenvalue problems
\begin{align}
  G_{ij}\rb{\tau_0 + \Delta\,\tau}\,u_{\alpha,j} = \expe{-E_\alpha\,\Delta\,\tau}\,G_{ij}\rb{\tau_0}\, u_{\alpha,j}, \\
  v_{\alpha,i}\, G_{ij}\rb{\tau_0 + \Delta\,\tau} = \expe{-E_\alpha\,\Delta\,\tau}\,v_{\alpha,i}\,G_{ij}\rb{\tau_0},
\end{align}
are solved for some suitable choice of pivot point $\rb{\tau_0, \,\Delta\,\tau}$ to obtain the eigenvectors $u_\alpha,\,v_\alpha$. Due to the symmetrisation of the matrix, these are identical. The effect of changing the pivot point was examined and found to be small. These eigenvectors are used to determine the projected correlators, each optimised for a single energy state: $G\rb{\tau,\alpha} = v_{\alpha,i}\,G_{ij}\rb{\tau}\,u_{\alpha,j}$.
\par
\begin{figure}[t]
  \centering

  \overlayText{0.21\textwidth}{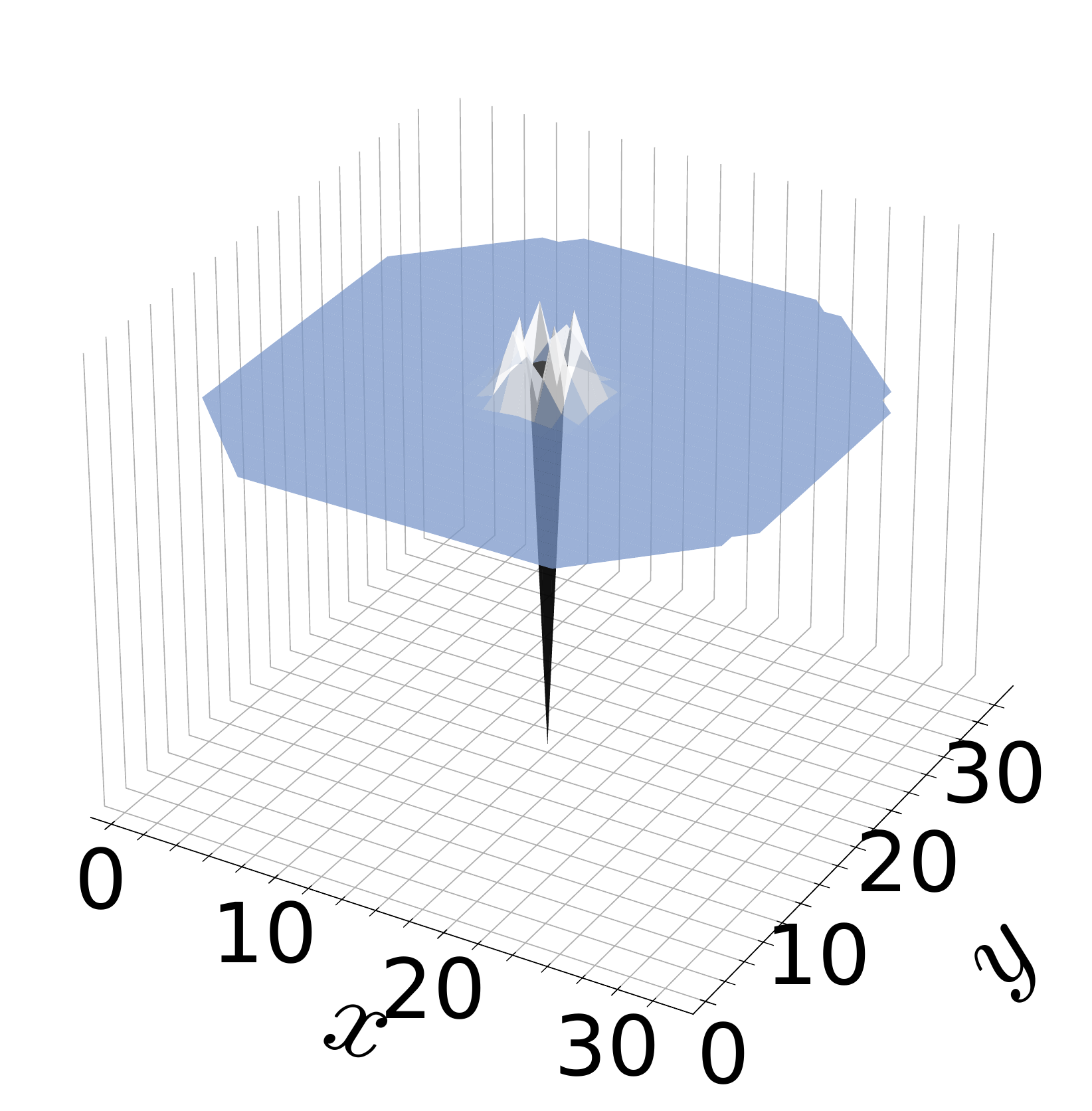}{0.05}{1.7}{$\sigma_E=1.0$}
  \overlayText{0.21\textwidth}{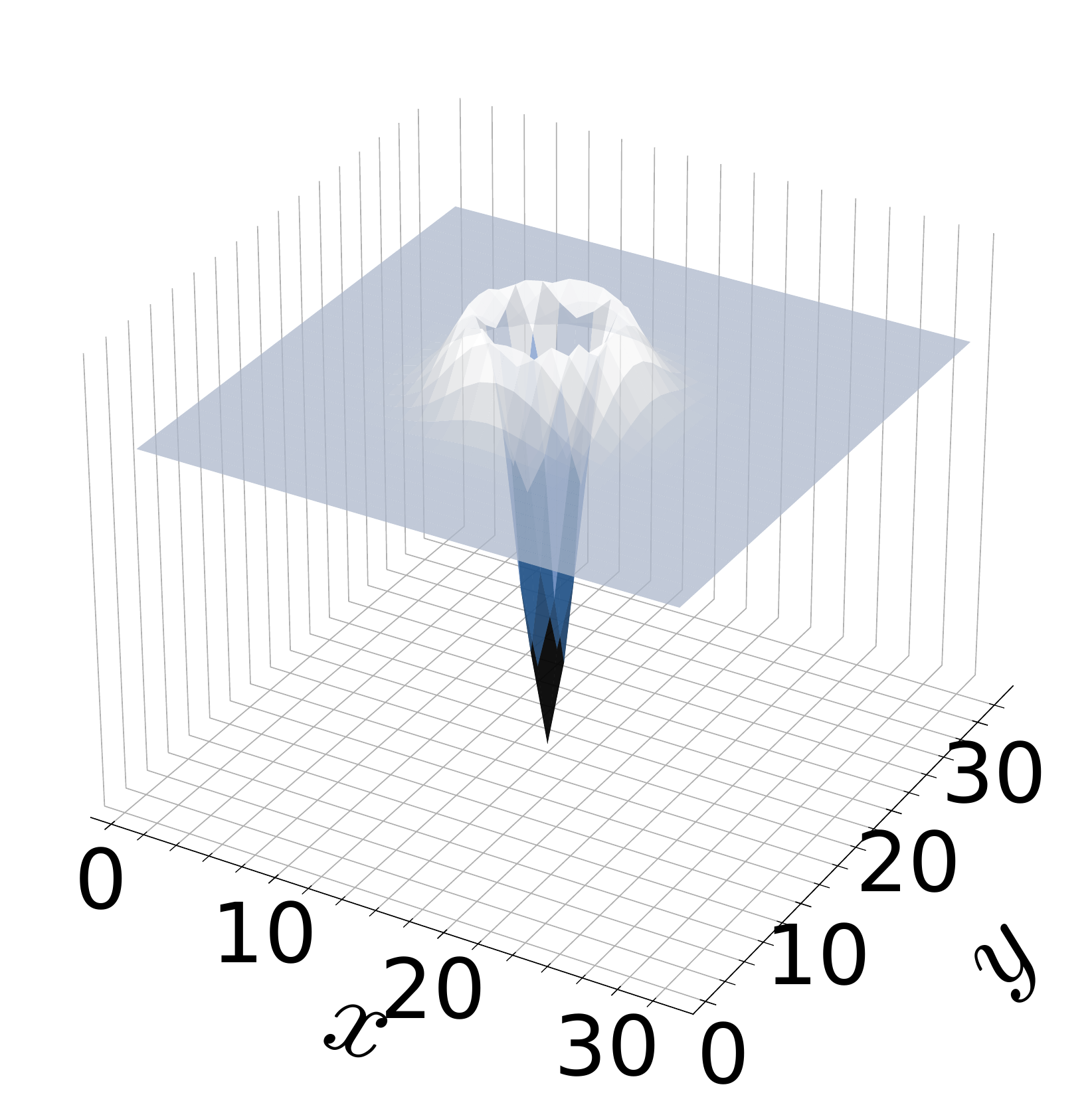}{0.05}{1.7}{$\sigma_E=2.5$}
  \overlayText{0.21\textwidth}{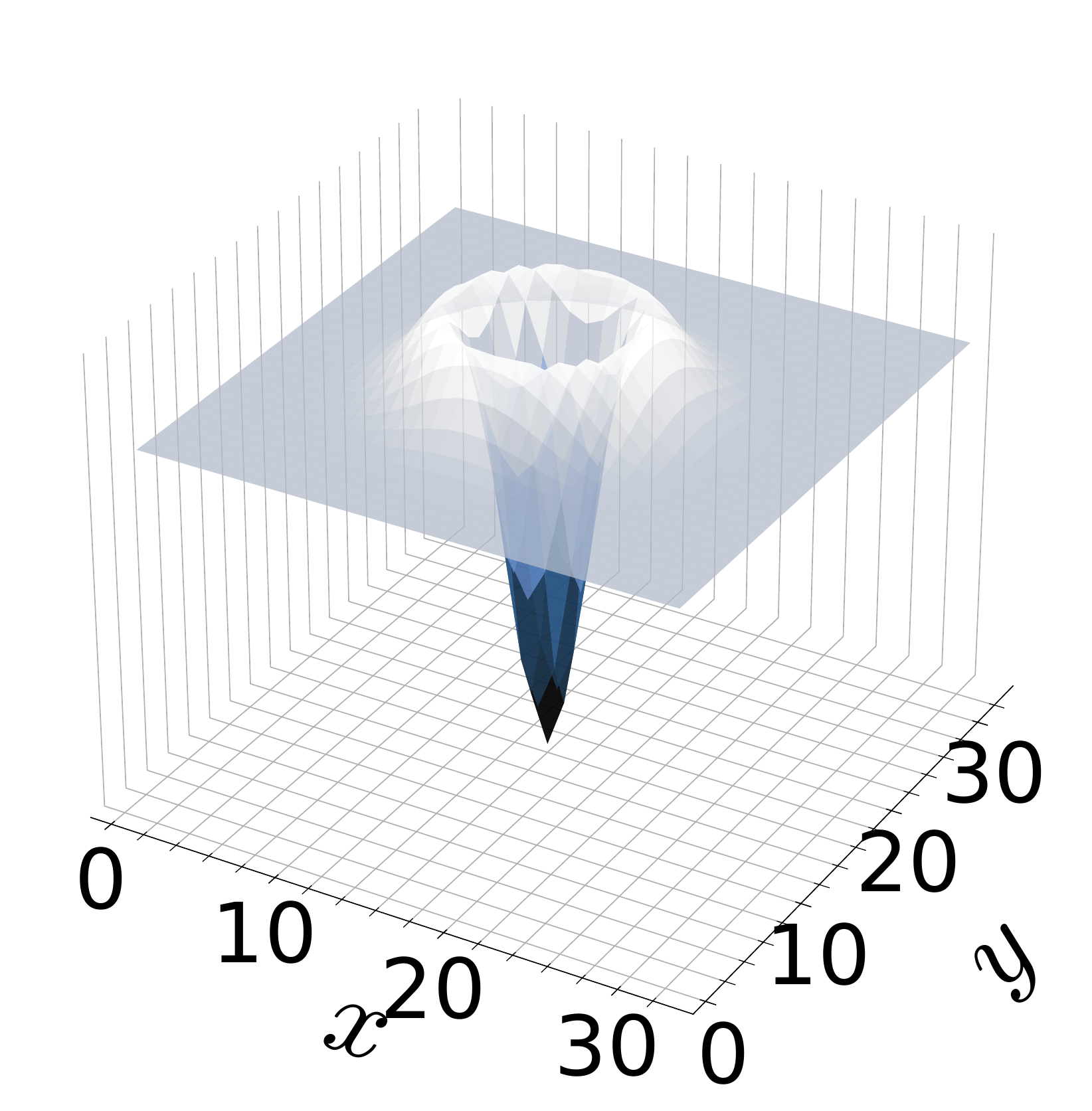}{0.05}{1.7}{$\sigma_E=3.5$}
  \overlayText{0.21\textwidth}{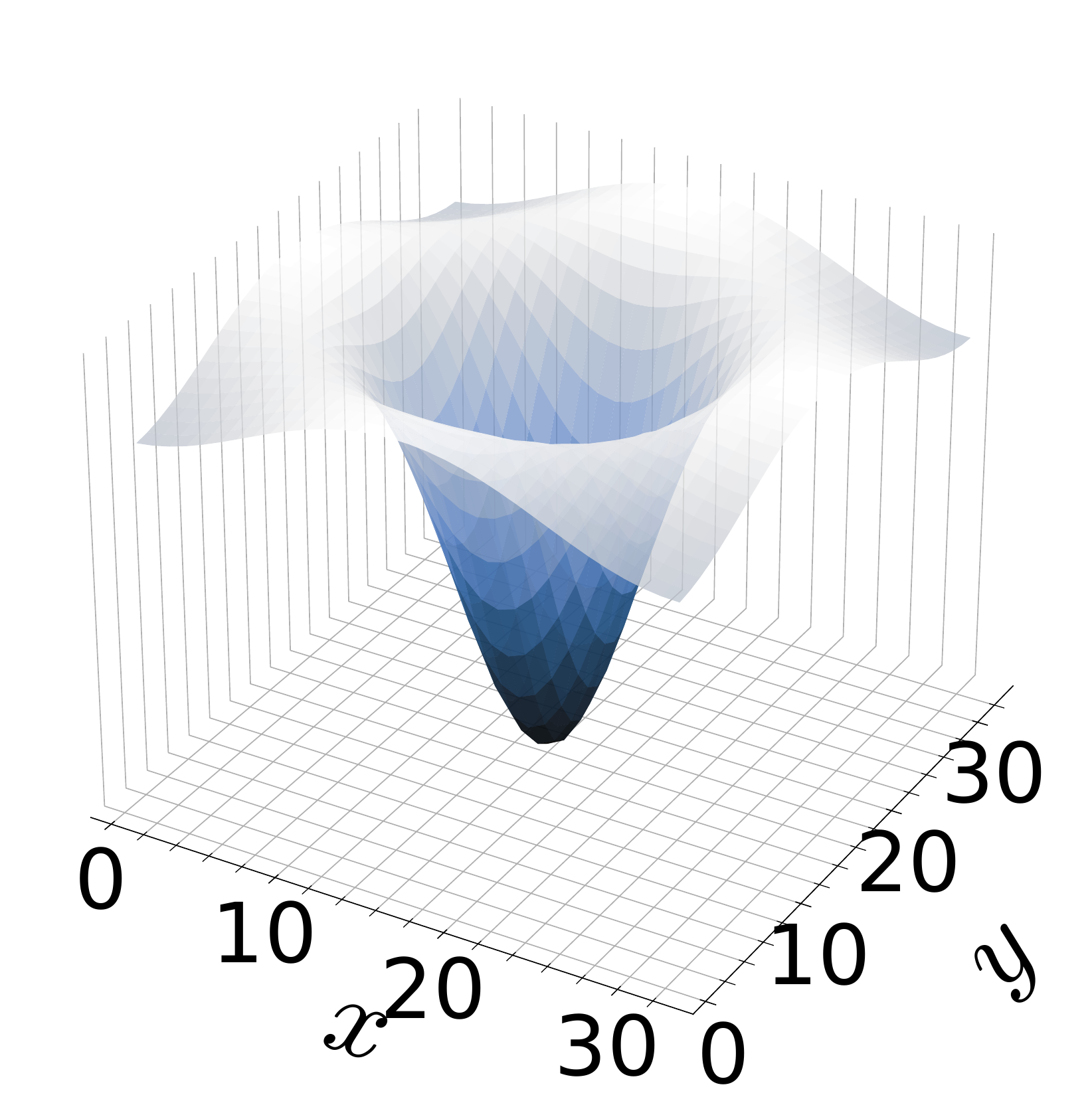}{0.05}{1.7}{$\sigma_E=8.0$}
  \\
  \overlayText{0.21\textwidth}{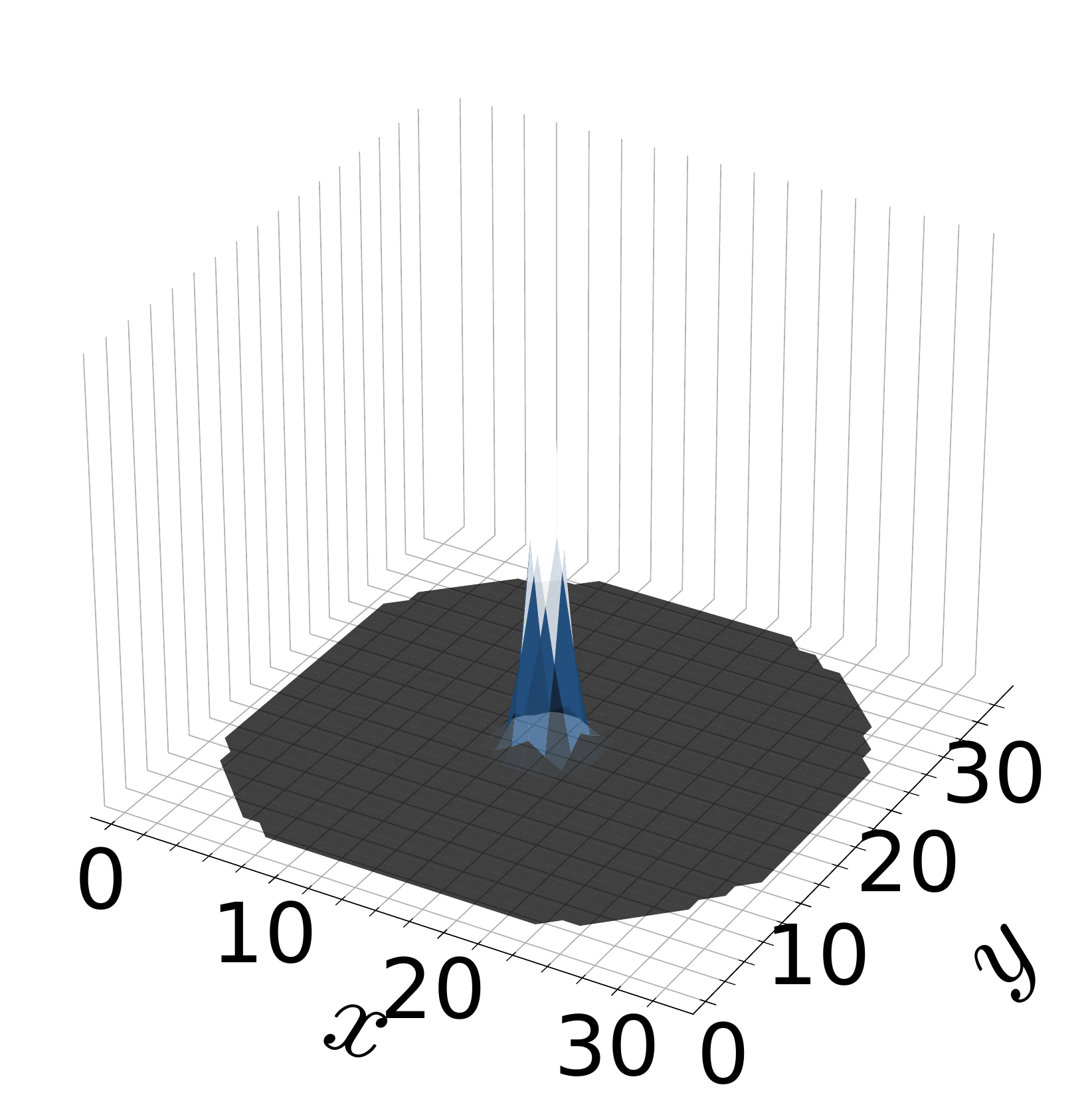}{0.05}{1.7}{$\sigma_S=1.0$}
  \overlayText{0.21\textwidth}{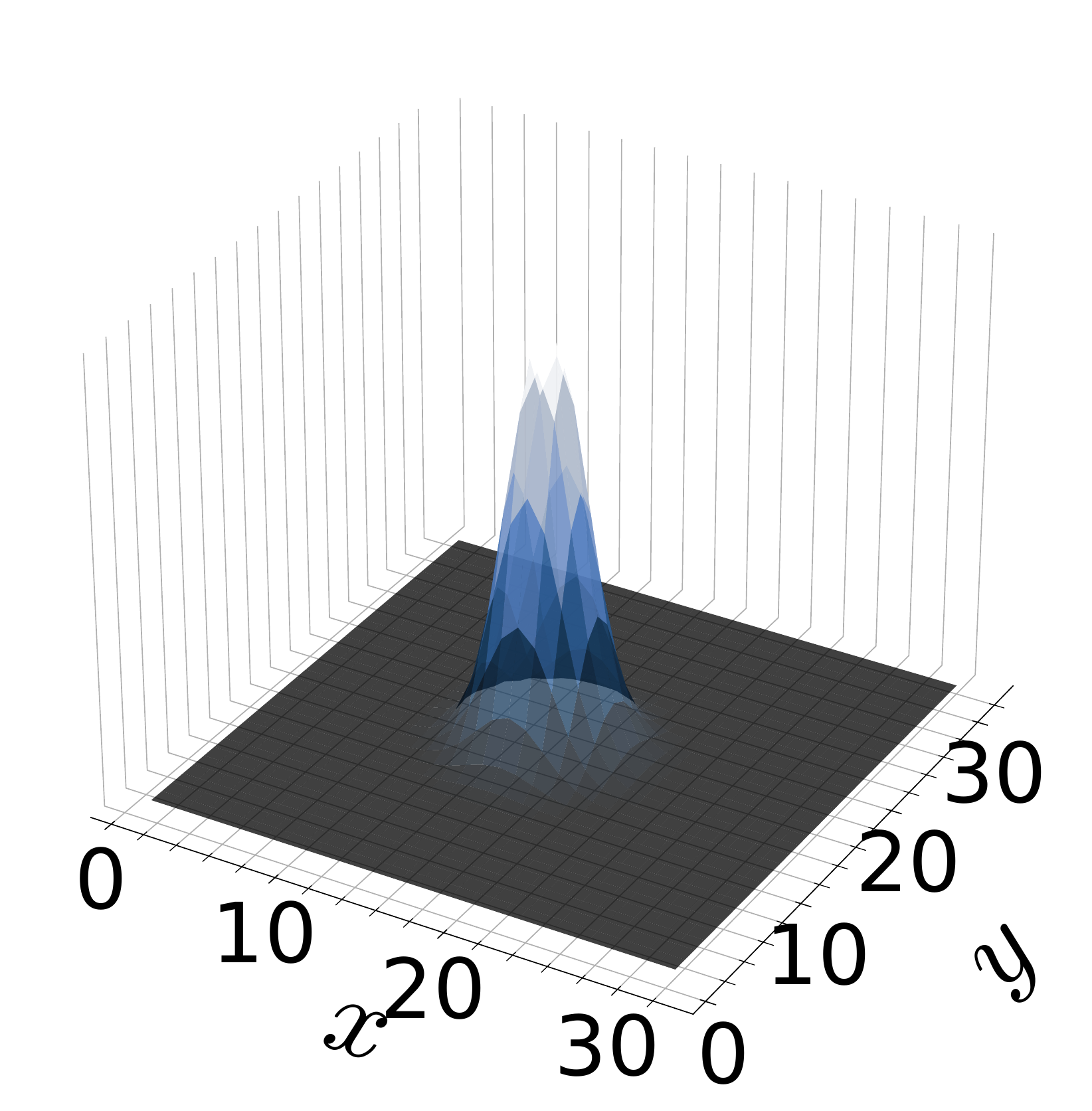}{0.05}{1.7}{$\sigma_S=2.5$}
  \overlayText{0.21\textwidth}{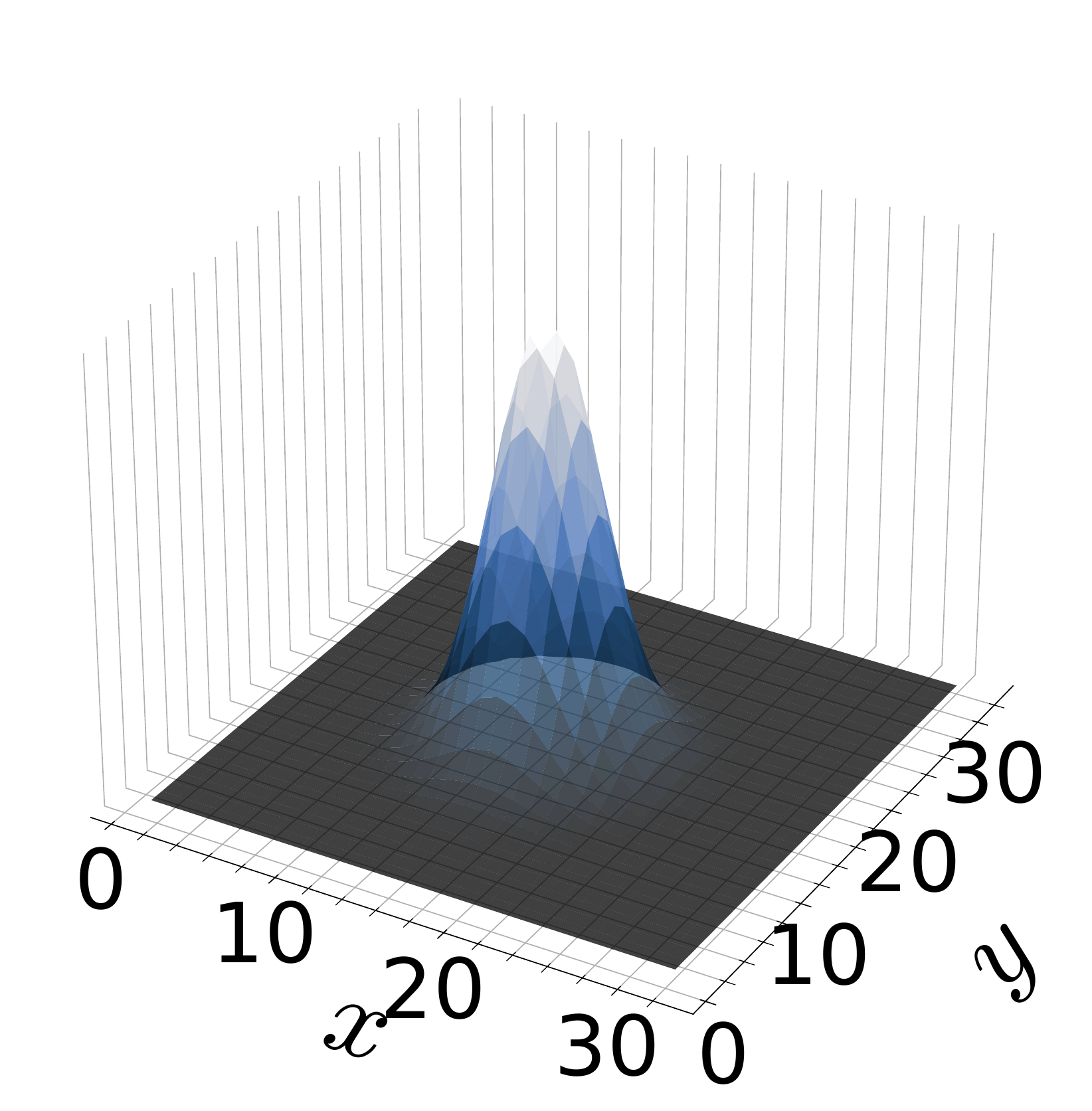}{0.05}{1.7}{$\sigma_S=3.5$}
  \overlayText{0.21\textwidth}{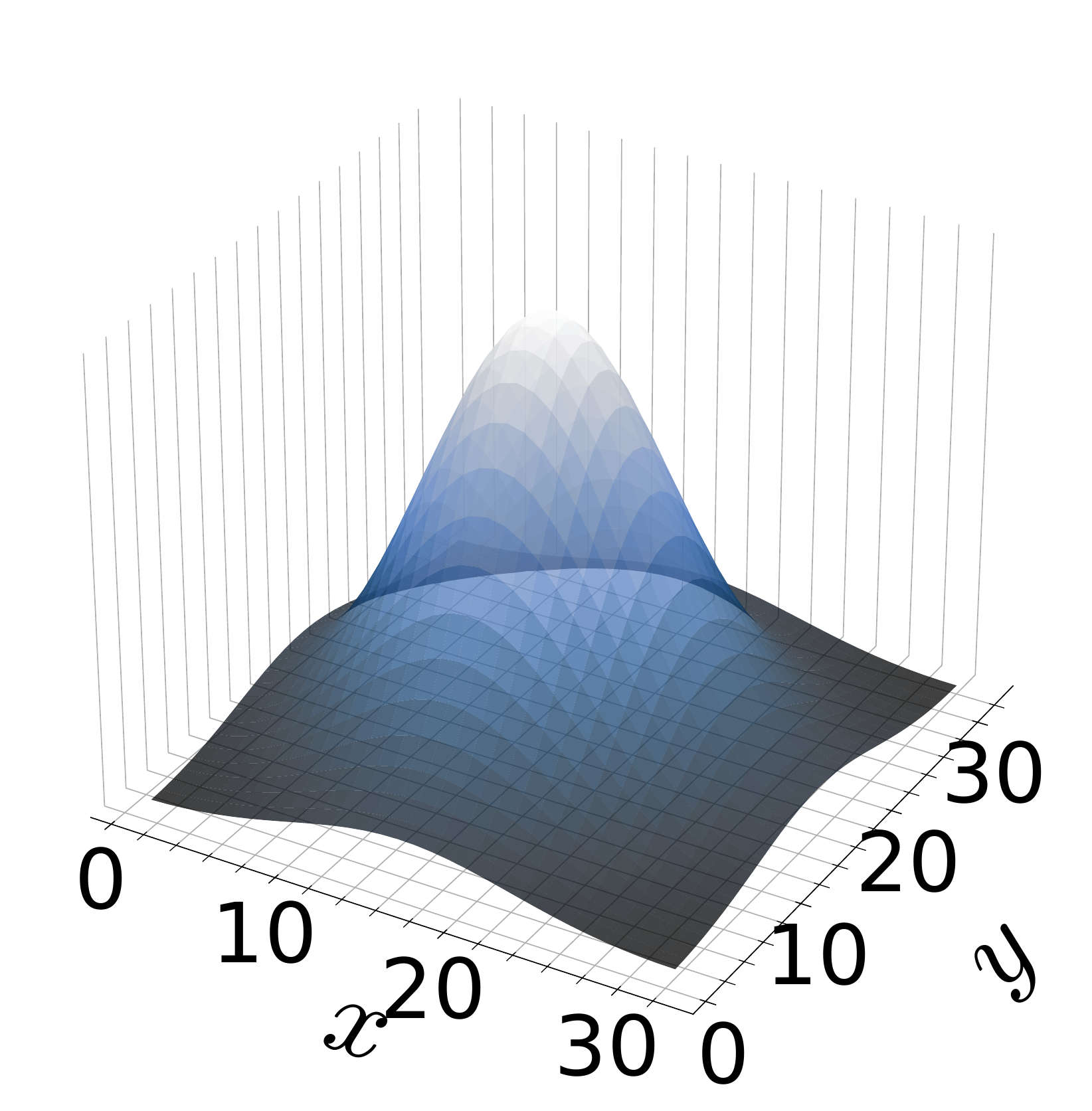}{0.05}{1.7}{$\sigma_S=8.0$}
  \caption{\tcr{Profiles} of the smeared (bottom) and \enquote{excited} (top) smearing kernels with four different \enquote{smearing} widths $\sigma = 1.0, \, 2.5, \,3.5,\, 8.0$. The \tcr{profiles} have been shifted such that the origin is at the centre of the $x-y$ plane. The overall normalisations in this figure are irrelevant. As the \enquote{smearing} widths are relatively small, periodic boundary conditions are not considered for these figures.}
  \label{fig:OP:smearings}
\end{figure}
In this study we make use of a set of Gaussian smearings and \enquote{excited} extended sources. For the Gaussians, the point source $\eta$ is smeared with a Gaussian profile
\begin{align}
      \eta_{S}\rb{x} = \rb{\frac{3}{2\,\pi\,\sigma^2}}^{\frac{3}{4}}\,\sum_{y}\,\expe{\frac{3\,\rb{x-y}^2}{4\,\sigma^2}}\,\eta\rb{y},
\end{align}
where $\sigma$ is the usual width parameter in a Gaussian of $\left<r^2\right> = \sigma^2$.
\par
For the \enquote{excited} sources the point source $\eta$ is smeared with the first radial excited wave function of the three-dimensional harmonic oscillator, in particular
\begin{align}
      \eta_{E}\rb{x} = \frac{1}{\sqrt{6}}\,\rb{\frac{3}{2\,\pi\,\sigma^2}}^{\frac{3}{4}}\,\sum_y\,\rb{\frac{3\,\rb{x-y}^2}{\sigma^2} -3}\,\expe{\frac{3\,\rb{x-y}^2}{4\,\sigma^2}}\,\eta\rb{y}.
\end{align}
This source has been observed to have a better overlap with $\rb{2S}$ excited states.
\par
In this study, we use four different \enquote{widths} for each of the Gaussian smeared (S) and excited (E) operators. These are $\sigma = 1.0, 2.5, 3.5, 8.0$. These widths were inspired by smearing radii which have previously been shown to be effective at isolating excited nucleon states~\cite{Mahbub:2010rm} and vary from narrow to very broad. This is evident in \Fig{fig:OP:smearings} where we plot the profiles of these operators.
\par
We find that the use of a correlation matrix comprising only the four Gaussian operators to be sufficient for all ground states considered but that an extended $8\times8$ matrix including the excited operators is particularly helpful for the higher excited states in the P-wave channel. This may be due to the different structure of the excited operators or may simply be due to the use of a larger basis of operators.
\section{Masses}
We follow the usual diagonalisation procedure, and the resulting \enquote{projected} correlators should have a large overlap with a single energy eigenstate. We fit this projected correlator with the appropriate form in order to extract the mass. At zero temperature where we expect the spectral function to be a sum of $\delta$-functions, this is easily done using a function of the form
\begin{align}
  G_{\text{fit}}\rb{\tau} = \sum_{i=1}^{N_{\text{exp}}}\,A_i\,\expe{-E_i\,\tau},
  \label{eqn:M:fit}
\end{align}
where we allow the number of exponential terms to vary between one and nine. As we use NRQCD, there is no backwards propagating state. We make use of model averaging techniques~\cite{Rinaldi:2019thf,NPLQCD:2020ozd,Jay:2020jkz} to incorporate knowledge from different fit windows. This approach resembles that presented in \Refl{Aarts:2023nax}. For the $\rb{3P}$ and \tcr{$\rb{4S}$} states we instead use standard constant plateau fits to the effective mass \tcr{as we find exponential fits unreliable for these noisy states}. 
\par
\begin{figure}[t]
  \centering
  \includegraphics[width=0.475\columnwidth]{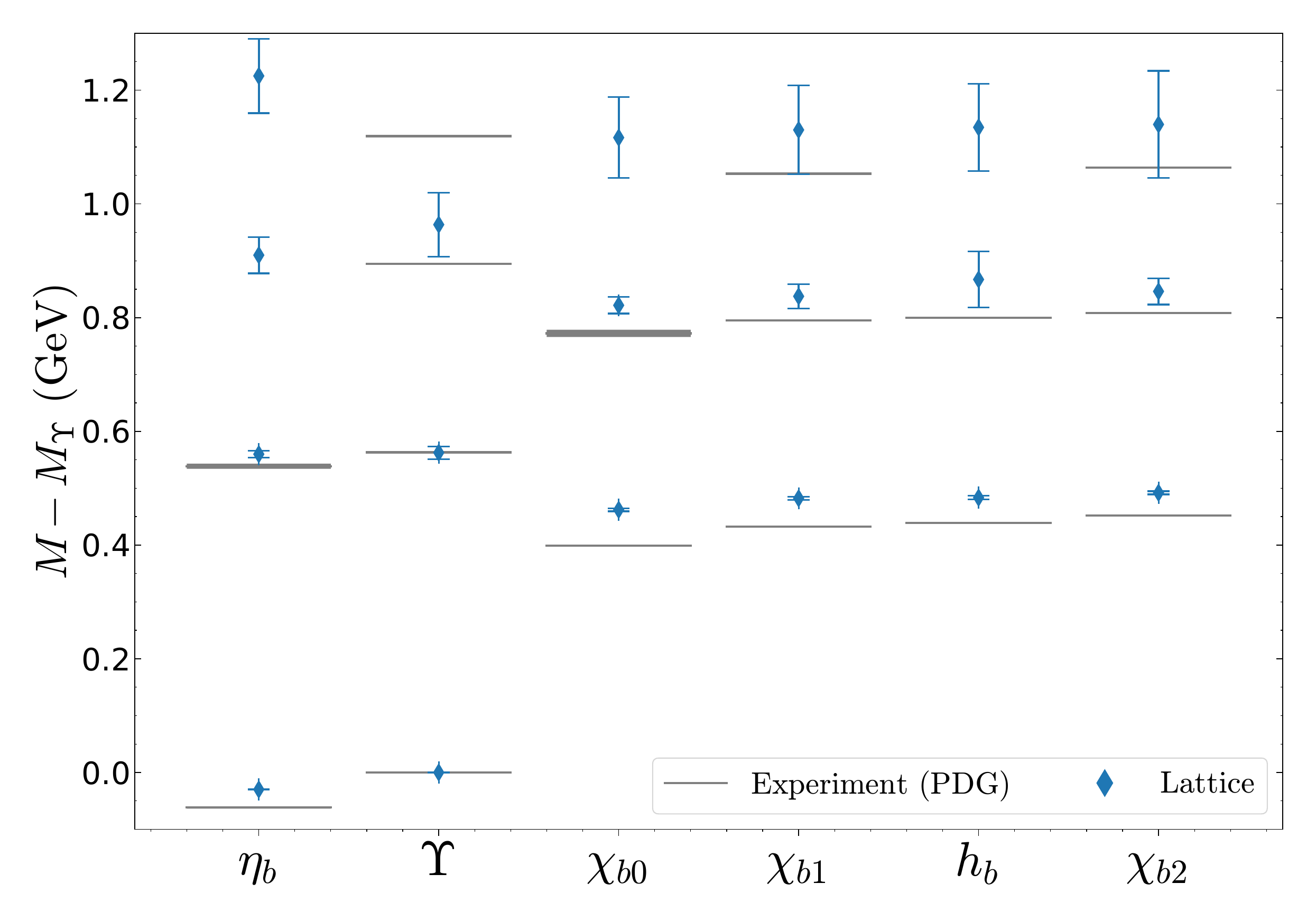}
  \includegraphics[width=0.475\columnwidth]{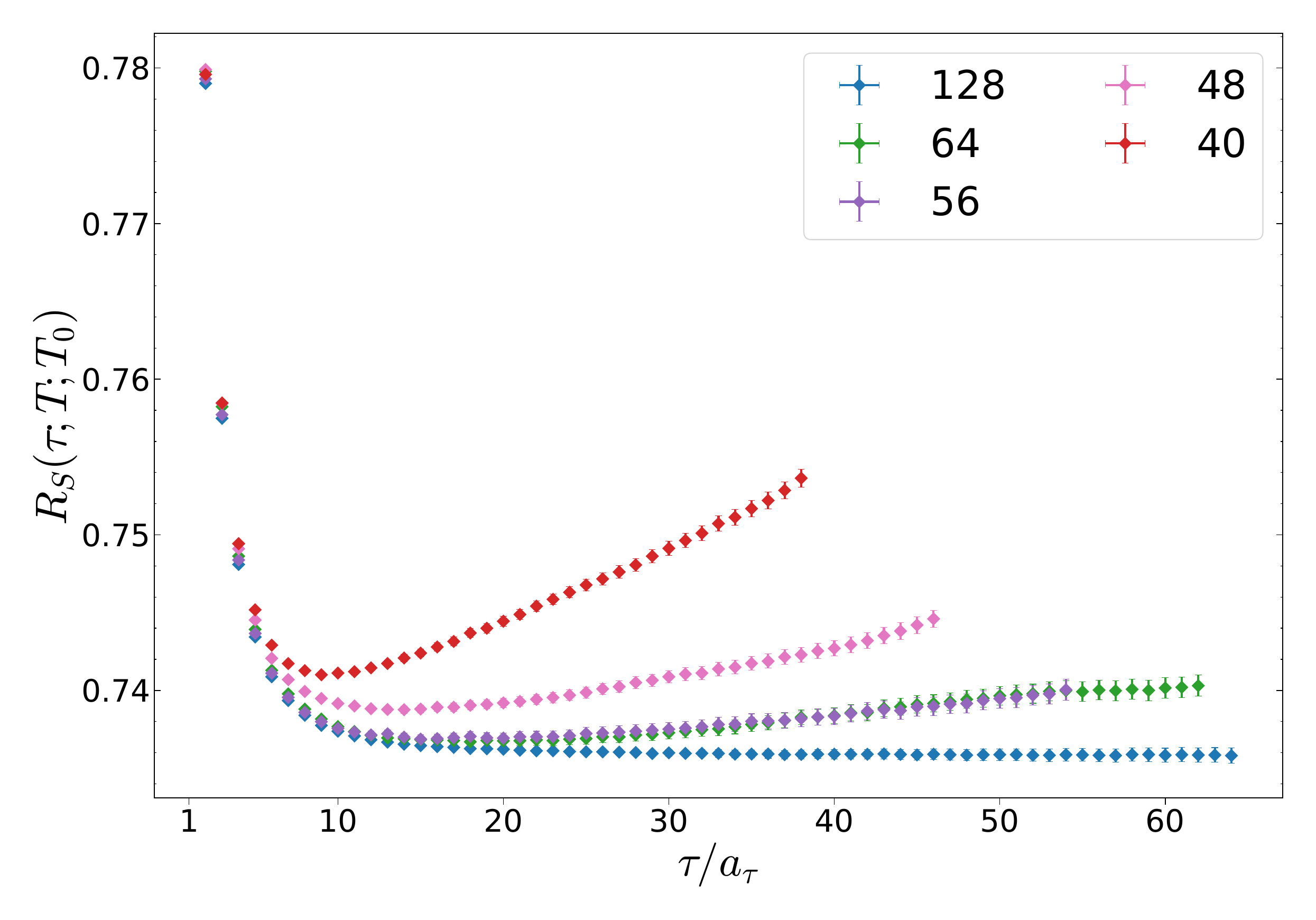}
  \\
  \includegraphics[width=0.475\columnwidth]{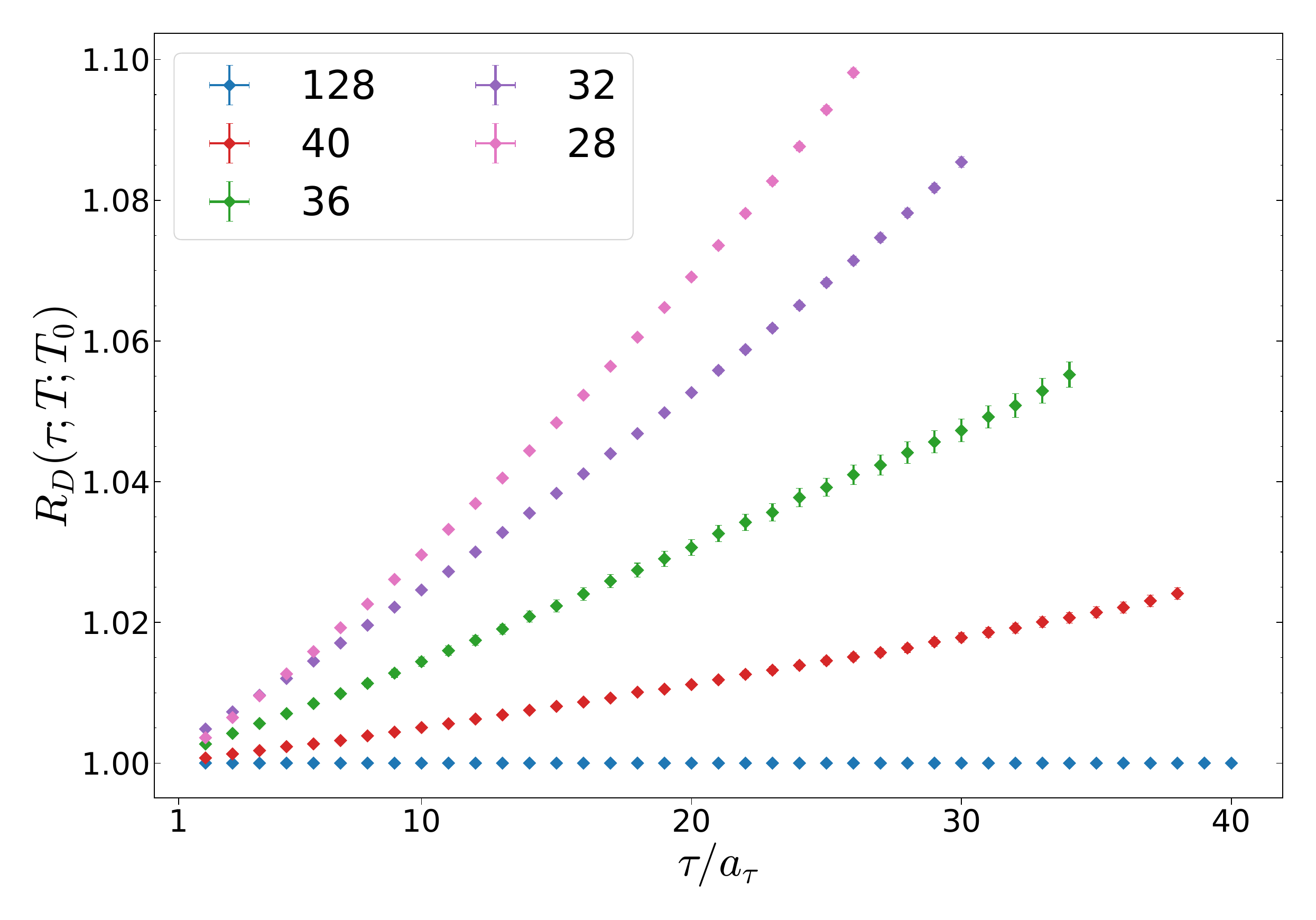}
  \includegraphics[width=0.475\columnwidth]{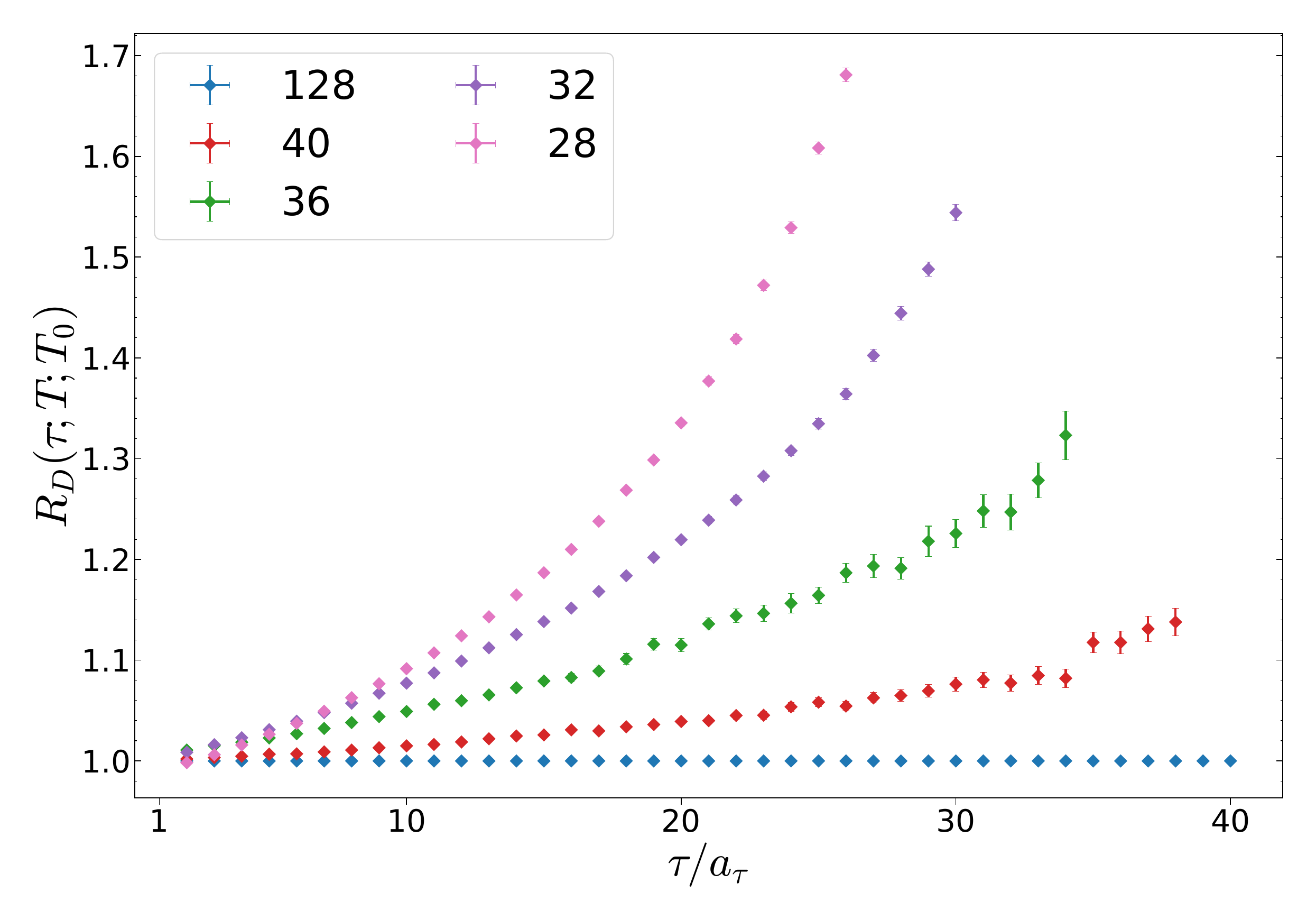}
  \caption{\textbf{Top Left:} Zero temperature bottomonia masses. The $\Upsilon(1S)$ mass has been subtracted off in each case as it is used to set the NRQCD additive mass shift. The experiment results are from the Particle Data Group~\cite{PhysRevD.110.030001}. \textbf{Top Right:} Single ratio $R_S$ of \eqnr{eqn:R:single} for the $\Upsilon(1S$). \textbf{Bottom Left (Right):} Double ratio $R_D$ of \eqnr{eqn:R:double} for the $\Upsilon(1S)$ and $(2S)$ respectively. Note that $N_\tau=128$ is equal to one by construction. These ratios show how much the spectral function is required to change from the one at zero temperature.}
  \label{fig:R:Ratio}
\end{figure}
The resulting zero-temperature masses are presented in the top left of \Fig{fig:R:Ratio}. Here the (lattice) zero-temperature $\Upsilon(1S)$ mass has been subtracted, thus removing the effect of the NRQCD additive mass shift. It is clear that in our simulation, the S-wave $\Upsilon$ states are well reproduced compared to their experimental values, but that the P-wave states $\text{ }\!\rb{\chi_{b0},\,\chi_{b1},\,\chi_{b2}, h_b}$ and the $\eta_{b}$ are systematically heavier. This is expected~\cite{HPQCD:2011qwj,Hudspith:2023loy} as we include only terms up to $\order{v^4}$ in the NRQCD expansion with only tree-level coefficients. As we are ultimately interested in the change as we increase the temperature, this is not a concern.
\subsection{Ratio Analysis}
The fit function of \eqnr{eqn:M:fit} is only appropriate in the case that the spectral function $\rho\rb{\omega}$ is a sum of well separated $\delta$-functions. In order to examine whether this is the case, without explicitly determining the spectral function, we turn to the single and double ratio analysis of \Refltwo{Aarts:2022krz}{Aarts:2023nax}. This approach is comparable to that of the reconstructed correlator ratio~\cite{Ding:2012sp,Kelly:2018hsi}.
\par
To investigate the change in the spectral function, we take a ratio of the lattice correlator to a model correlator that assumes a single $\delta$-function state with parameters determined at zero temperature $\rb{N_\tau = 128}$
\begin{align}
  R_S\rb{\tau;T,T_0} = \frac{G\rb{\tau;T}}{G_{\text{model}}\rb{\tau;T,T_0}} = \frac{G\rb{\tau;T}}{A_0\rb{T_0}\,\expe{-E_0\rb{T_0}\,\tau}}.
  \label{eqn:R:single}
\end{align}
If the spectral function of the lattice correlator $G\rb{\tau;T}$ contains only a single $\delta$-function, this ratio will be a constant. It is clear from \Fig{fig:R:Ratio} (top right) that this is not the case. One part of this is the presence of excited states in the lattice correlator but not the model and hence we turn to the double ratio
\begin{align}
  R_D\rb{\tau;T,T_0} = \frac{R_S\rb{\tau;T,T_0}}{R_S\rb{\tau;T_0;T_0}},
  \label{eqn:R:double}
\end{align}
where we take a ratio of the single ratio at finite temperature $T$ and zero temperature $T_0$. In this ratio, the model correlators $G_{\text{model}}$ exactly cancel as the NRQCD kernel $K\rb{\tau,\omega} = \exp\rb{-\omega\,\tau}$ has no temperature dependence~\cite{Aarts:2014cda}. This ratio acts to remove the effect of excited states if they are the same at $T$ and $T_0$. Any change in the finite temperature spectral function is now shown by deviations away from one. This is shown in the bottom of \Fig{fig:R:Ratio} (left) for the $\Upsilon(1S)$ and (right) for the $\Upsilon(2S)$. It is clear that temperature effects appear immediately, even at small $\tau$ for these temperatures.
\par
\begin{figure}[t]
  \centering
  \begin{tikzpicture}
    \centering
    \node at (0, 0) {\includegraphics[width=0.475\columnwidth, keepaspectratio,origin=c]{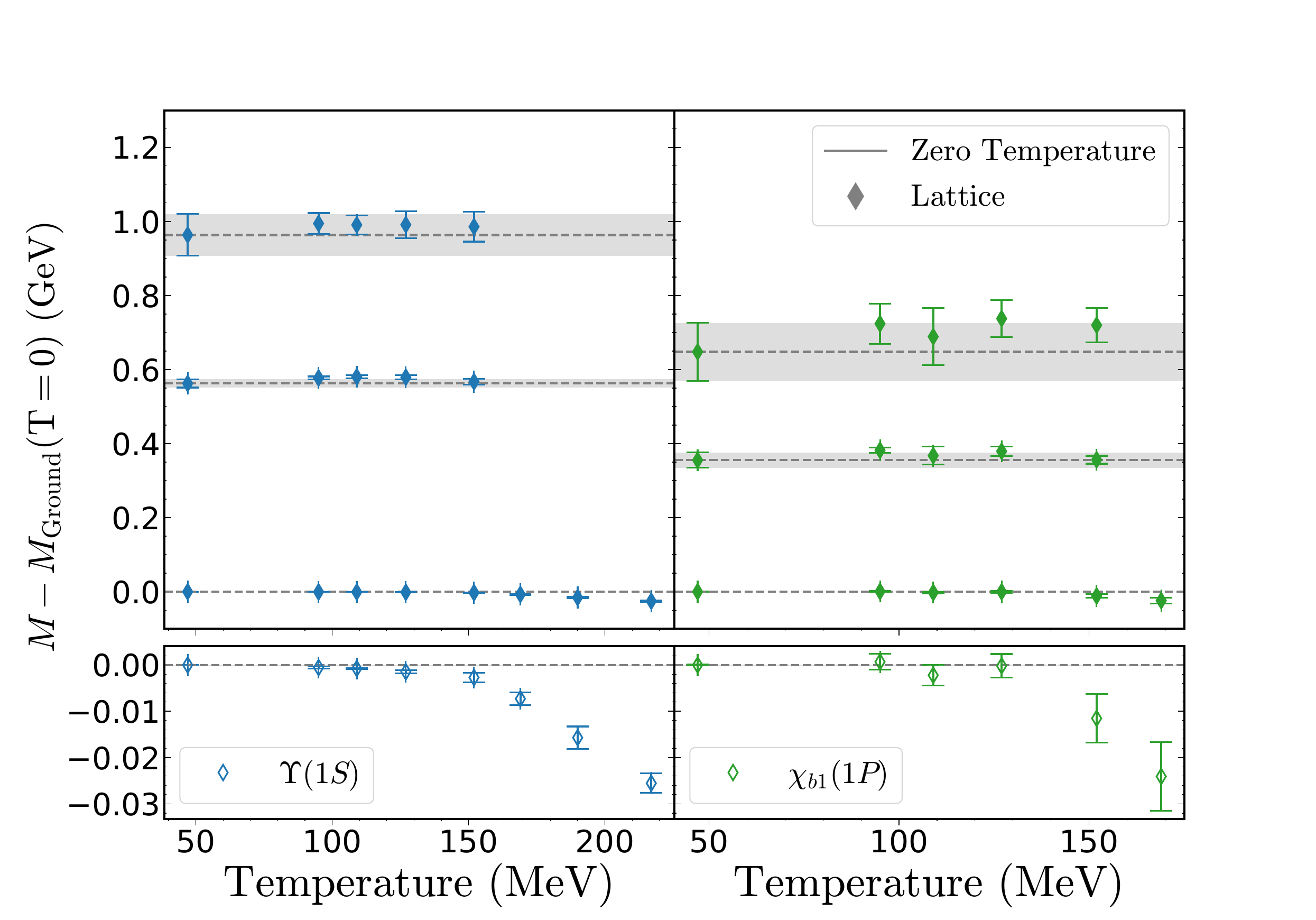}};
    \node at (-2.3, -0.5) {\tiny$\Upsilon\rb{1S}$};
    \node at (0.9, -0.5) {\tiny$\chi_{b1}\rb{1P}$};
    \node at (-2.3, 0.6) {\tiny$\Upsilon\rb{2S}$};
    \node at (0.9, 0.25) {\tiny$\chi_{b1}\rb{2P}$};
    \node at (-2.3, 1.5) {\tiny$\Upsilon\rb{3S}$};
    \node at (0.9, 0.95) {\tiny$\chi_{b1}\rb{3P}$};
    \end{tikzpicture}
  \includegraphics[width=0.475\columnwidth,page=4]{Figures/PlotV_AV_mom.pdf}
  \caption{\textbf{Left: }Zero temperature bottomonia masses. The zero-temperature ground state mass has been subtracted off in each channel. A close up of each ground states is shown in the lower of figure. \textbf{Right: } Width extracted from the time-derivative method for ground and first excited $\Upsilon$ and $\chi_{b1}$ states as a function of temperature.}
  \label{fig:R:Mass}
\end{figure}
It is clear that the $\Upsilon(1S)$ is much less affected by temperature than the $\Upsilon(2S)$ and so we may consider fitting using the standard exponential fit function of \eqnr{eqn:M:fit} to a higher temperature than for the $\Upsilon(2S)$. We repeat this examination as in \Refl{Aarts:2023nax} for all temperatures \tcr{and $\Upsilon$ and $\chi_{b1}$ states. Where possible, we show the }zero-temperature subtracted mass results in \Fig{fig:R:Mass} (left).
\par
The $\Upsilon(1S)$ and $\chi_{b1}\rb{1P}$ show a slight decrease in mass as the temperature increases, with the ratio analysis suggesting that it is possible to extract a mass well past the pseudocritical temperature of $T_{c} \sim 167$ MeV for the $\Upsilon\rb{1S}$. The $\Upsilon$ excited states and the $\chi_{b1}$ states are noisier and the ratio analysis suggests a greater change to the spectral function as the temperature increases. 

\section{Time-Derivative Moments}
The recently introduced time-derivative moments method~\cite{RachelProceedings} is well suited to examine the ground state properties of a given correlator. As such, when applied to the projected correlators produced by the GEVP above, it will tell us the excited state properties. Here we will briefly recap the main features of the method as it relates to the width of the spectral function
\par
Begin by assuming that the spectral function is comprised of a sum of Gaussians with means $E_i$ and widths $\Gamma_i$ so that we can write
\begin{align}
  G\rb{\tau} &= \sum_{i=0}^{\infty} \, A_i\,\exp\rb{-\tau\sq{E_i - \frac{\Gamma^2_i\,\tau}{2}}} \nonumber\\
  &= A_0\,\exp\rb{-\tau\sq{E_0 - \frac{\Gamma^2_0\,\tau}{2}}}\,\sq{1 + \sum_{i=1}^{\infty} \, \frac{A_i}{A_0}\,\exp\rb{-\tau\sq{\rb{E_i -E_0} - \frac{\rb{\Gamma^2_i-\Gamma^2_0}\,\tau}{2}}}},
\end{align}
where $A_i$ is related to the overlap of each state to the operators used. The first log-derivative of this is then (using the $\log(1+x)\approx x $ for small $x$ approximation)
\begin{align}
      \frac{\partial\log{G(\tau)}}{\partial\tau} = \rb{-E_0 + \Gamma_0^2\tau} + \sum_{i=1}^{\infty}\frac{A_i}{A_0}\rb{-\tau\rb{\rb{E_i-E_0} - \rb{\Gamma_i^2-\Gamma_0^2}\tau}}\expe{-\tau\sq{\rb{E_i -E_0} - \frac{\rb{\Gamma^2_i-\Gamma^2_0}\tau}{2}}}
\end{align}
and the second is
\begin{align}
  \frac{\partial^2\log{G(\tau)}}{\partial\tau^2} = \Gamma^2_0 + \sum_{i=1}^{\infty}\frac{A_i}{A_0}\left[\rb{-\rb{E_i-E_0} + \rb{\Gamma_i^2-\Gamma_0^2}\tau}^2 + \rb{\Gamma_i^2-\Gamma_0^2}\right]\expe{-\tau\sq{\rb{E_i -E_0} - \frac{\rb{\Gamma^2_i-\Gamma^2_0}\tau}{2}}}
  \label{eqn:Mo:log2}
\end{align}
As we are interested in the width of the state as the temperature changes, we will construct the second log-derivative of \eqnr{eqn:Mo:log2} using a fourth-order finite difference operator using the \textsc{FinDiff}~\cite{findiff} \textsc{python} package.
{
A fourth-order method was found to be a sensible compromise between accuracy and the number of points used (particularly relevant at shorter, hotter temperatures). We will then fit with the function
\begin{align}
    \frac{\partial^2\,\log{G(\tau)}}{\partial\,\tau^2} = \Gamma^2 + \sum_{i=1}^{N}\,a_i\exp\rb{-b_i\,\tau + \frac{\rb{c_i^2 - \Gamma^2}}{2}\,\tau^2},
\end{align}
which assumes that $\rb{E_i-E_0} \gg \rb{\Gamma_i^2-\Gamma_0^2}\tau$. In practice, we find little difference between fits with or without the $\rb{c_i - \Gamma^2}\,\tau^2$ term as our projected correlators have reduced overlap with excited states. This term is additionally exponentially suppressed by the additional factor of $\tau$. The only term that we care about here is the $\Gamma^2$ term pertaining to the width of the state. In this manner, we are robust against time-dependence and may be able to extract a clean signal given otherwise noisy data.
Let's now apply this method at a single, common fit window of $[4, N_\tau -4]$ to the GEVP projected correlators. The results are shown in \Fig{fig:R:Mass} (right). There is a clear hierarchy in the observed widths: excited states are broader than ground states and the $\chi_{b1}$ is broader than the $\Upsilon$. This is in line with previous determinations for the ground states and our expectations for the excited states~\cite{Mocsy:2008eg,Aarts:2011sm,Larsen:2019zqv,Skullerud:2022yjr,Spriggs:2021dsb,Aarts:2023nax,Ding:2025fvo}.
\section{Summary}
In this work, the temperature dependence of both the mass and the width -- through the time derivative moments approach -- of the ground and excited state $\Upsilon$ and $\chi_{b1}$ have been elucidated. We find that there is \tcr{an indication of a} negative shift in the mass of the $\Upsilon\rb{1S}$ and the $\chi_{b1}\rb{1P}$ but that the uncertainties for the other $\Upsilon$ and $\chi_{b1}$ states investigated are compatible with no change. The widths of all the states are extracted and it is found at all temperatures that the widths follow the hierarchy $\chi_{b1}(2P) > \Upsilon(2S) > \chi_{b1}(1P) > \Upsilon(1S)$. Particularly of note is the slow change of the $\Upsilon(1S)$ past the pseudocritical temperature while the other states change much faster past $T_c$.
\par
The excited states in the study were obtained by solving a generalised eigenvalue problem using a matrix of generic smeared operators. This is an approach common in zero-temperature studies and less so at finite temperature~\cite{Larsen:2019zqv,Ding:2025fvo}. It is found that the Gaussian smeared operators are well able to project the different states with only minimal impact from the inclusion of the node-like \enquote{excited} operators.
\par
In the future we plan to explore further use of GEVP projected correlators alongside other methods of spectral function extraction for excited state properties, examine the effect of the number of data points at each temperature via a set of ensembles with increased anisotropy~\cite{Skullerud:2022yjr} and further increase the statistics used in this study.
\appendix
\section{Software \& Data}
\Fig{fig:OP:smearings} uses a perceptually uniform colour map~\cite{crameri2020misuse} available from \Refltwo{crameri_2023_8409685}{github_cmcrameri}. This analysis makes extensive use of the \textsc{python} packages \textsc{gvar}~\cite{peter_lepage_2024_12675777} and \textsc{lsqfit}~\cite{peter_lepage_2024_12690493}. The \enquote{Time-Derivative Moments} analysis uses the \textsc{FinDiff}~\cite{findiff} \textsc{python} package. Additional data analysis tools included \textsc{matplotlib}~\cite{Hunter:2007,thomas_a_caswell_2022_6513224} and \textsc{NumPy}~\cite{harris2020array}. Error analysis is performed through a combination of \textsc{gvar} and a jackknife analysis~\cite{Efron1979} implemented in \textsc{Fortran} using the \textsc{Fortran-Package-Manager}~\cite{DBLP:journals/corr/abs-2109-07382,779aad0a0cba4c0297f31b532bd4aca7} with \textsc{python} bindings~\cite{fortran_meson}. The NRQCD correlators were produced using the package available from \Refl{FASTNRQCD}.

\acknowledgments
We acknowledge EuroHPC Joint Undertaking for awarding the project EHPC-EXT-2023E01-010 access to LUMI-C, Finland. This work used the DiRAC Data Intensive service (DIaL2 \& DIaL) at the University of Leicester, managed by the University of Leicester Research Computing Service on behalf of the STFC DiRAC HPC Facility (www.dirac.ac.uk). The DiRAC service at Leicester was funded by BEIS, UKRI and STFC capital funding and STFC operations grants. This work used the DiRAC Extreme Scaling service (Tesseract) at the University of Edinburgh, managed by the Edinburgh Parallel Computing Centre on behalf of the STFC DiRAC HPC Facility (www.dirac.ac.uk). The DiRAC service at Edinburgh was funded by BEIS, UKRI and STFC capital funding and STFC operations grants. DiRAC is part of the UKRI Digital Research Infrastructure. This work was performed using the PRACE Marconi-KNL resources hosted by CINECA, Italy. We acknowledge the support of the Supercomputing Wales project, which is part-funded by the European Regional Development Fund (ERDF) via Welsh Government. This work is supported by STFC grant ST/X000648/1 and The Royal Society Newton International Fellowship. RB acknowledges support from a Science Foundation Ireland Frontiers for the Future Project award with grant number SFI-21/FFP-P/10186. RHD acknowledges support from Taighde Éireann – Research Ireland under Grant number GOIPG/2024/3507. We are grateful to the Hadron Spectrum Collaboration for the use of their zero temperature ensemble. 
\FloatBarrier
\bibliographystyle{JHEP}
\bibliography{skeleton}

\end{document}